# Debye temperature of single-crystal Cr with incommensurate and commensurate magnetic structure

Stanisław M. Dubiel[1*], Jan Żukrowski[2]

AGH University of Science and Technology, [1]Faculty of Physics and Applied Computer Science, [2]Academic Center for Materials and Nanotechnology, PL-30-059 Kraków, Poland

**Abstract**

Single-crystals of Cr and Cr3%Mn doped with ~0.2 at.% $^{119}$Sn were studied by means of $^{119}$Sn-site Mössbauer spectroscopy. Mössbauer spectra recorded in the temperature range of 78-300 K were analyzed using a hyperfine field distribution method. Derived therefrom values of a center shift, CS, analyzed in terms of the Debye model yielded values of the Debye temperature, $T_D = 477(56)$K for Cr and $T_D = 282(16)$K for Cr3%Mn. The difference in the $T_D$-value can be regarded as evidence that a spin-phonon coupling is different for the incommensurate and commensurate antiferromagnetic structure of chromium.

Metallic chromium is regarded as an archetype of a antiferromagnet (AF) with the Nèel temperature $T_N \approx 313$ K [1]. However, it is not a normal AF as magnetic moments, $\mu$, have no constant magnitude but are harmonically modulated in space i.e. $\mu = \mu_1 sin\alpha + \mu_3 sin3\alpha$, where $\alpha = \mathbf{q} \cdot \mathbf{r}$, $\mathbf{q}$ is a wave vector and $\mathbf{r}$ is a distance in the real space, $\mu_1$ is an amplitude of the first-order harmonic and $\mu_3$ the one of the third-order. The ratio $\mu_3 / \mu_1$ ranges between 1.5 and 2.5 % depending on temperature [2,3]. The harmonic modulation of $\mu$ in real space gives rise to name the effect spin-density waves (SDWs). There are two types of these SDWs viz. (a) longitudinally polarized ones with $\mathbf{q} \parallel \mathbf{\mu}$, LSDWs, and (b) transversely polarized, TSDWs, with $\mathbf{q} \perp \mathbf{\mu}$. The former exist below the so-called spin-flip temperature, $T_{SF}=123$ K, and the latter between $T_{SF}$ and $T_N$. SDWs are related to a topology of the Fermi surface and are very sensitive to foreign atoms. Especially strong effect on the SDWs have neighbor atoms i.e. vanadium and manganese. Vanadium, having one valence electron less acts as an electron acceptor and its addition into Cr has a destructive effect on the SDWs. Indeed, addition of only ~4 at.% V completely destroys the SDWs [3,4]. On the other hand, Mn with one electron more acts as an electron donor, so its addition into Cr supports the SDWs [4]. In fact, on addition of Mn increases the value of $T_N$ and the amplitude of $\mathbf{\mu}$ but also rapidly grows the value of the wave vector, $\mathbf{q}$. The latter causes a change of the character of the SDWs: when the concentration of Mn $\geq$ 1 at.% the magnetic structure of Cr can be termed as a normal AF. The basic method for studying magnetic properties of Cr is diffraction of neutrons. Using this technique all AF features of chromium mentioned above were revealed. However, Mössbauer spectroscopy (MS) using the effect on [119]Sn atoms has turned out to be very useful with this respect. In particular, [119]Sn spectra yielded correct information on the Nèel temperature [6], an amplitude and sign of the 3[rd]-order harmonics [2], spin-flip transition [7]. This means that non-magnetic Sn atoms do not disturb the SDWs hence they can be used as the adequate probe to study properties of the SDWs in Cr. An advantage of applying MS is a possibility of getting information on lattice dynamics, hence prospectively to study an effect of magnetism on lattice vibrations. There is a general believe that magnetism has negligible effect on these vibrations because a contribution of an electron-phonon interaction (EPI) to magnetization of metallic systems is expected to be small, as, in general $E_D/E_F \leq 10^{-2}$ [8], where $E_D$ is the Debye energy and $E_F$ is the Fermi one. However, according to Kim [8] the effect of the EPI can be strongly enhanced below the Curie temperature, $T_C$, in an itinerant ferromagnet. Indeed, our recent studies performed with MS on topologically close-packed (TCP) compounds sigma-phase $Fe_{60}V_{40}$ [9] and C14 Laves phase $NbFe_2$ [10] revealed significant disturbance of the lattice vibrations on going from a paramagnetic into ferromagnetic state. The aim of the present study was to see whether or not the lattice dynamics of Cr with incommensurate SDWs is the same as the one of Cr with normal AF structure (commensurate). For this purpose we recorded a set of [119]Sn-site Mössbauer spectra in the temperature range of 78-300 K on two single-crystal samples viz. (a) Cr and (b) Cr3%Mn. Both samples were doped with ~0.2 at.% [119]Sn with a procedure described elsewhere [2]. According to the magnetic phase diagram of Cr-Mn [4], the (b) sample should be in the normal AF state with $T_N \approx 550$ K. Undeniably, the shape of the spectrum recorded on Cr3%Mn and shown in Fig. 1 is in line with this expectation, while the spectrum recorded on Cr has features characteristic of the incommensurate SDWs [2].

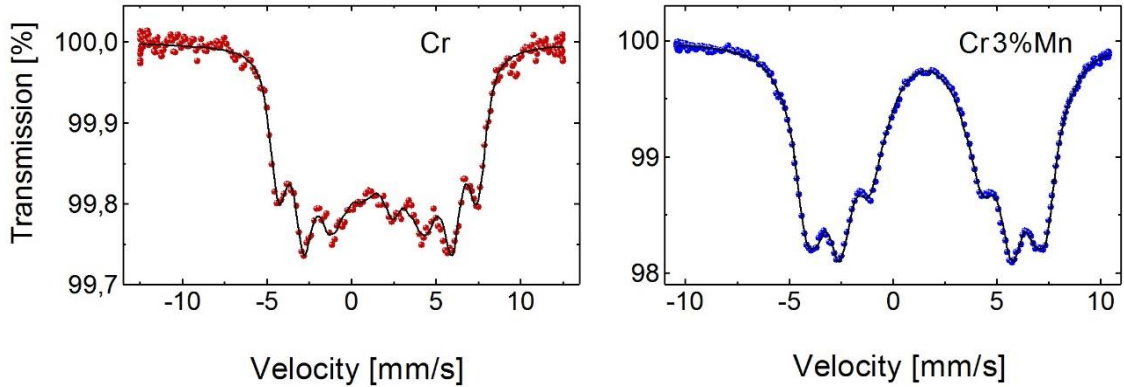

Fig. 1
[119]Sn spectra recorded at 80 K on single-crystals of Cr (left) and of Cr3%Mn (right) doped with ~0.2 at% [119]Sn.

There are two spectral parameters that are relevant to the lattice dynamics i.e. (1) center shift, CS, and (2) spectral area. Here we will consider only the former as it can be determined with much better accuracy than the latter.

A temperature dependence of the center shift, *CS(T),* can be expressed by the following equation:

$$CS(T) = IS(0) - \frac{3k_B T}{2mc}\left[\frac{3T_D}{8T} + \left(\frac{T}{T_D}\right)^3 \int_0^{T_D/T} \frac{x^3}{e^x - 1}dx\right]$$

Where *IS(0)* stays for the isomer shift (temperature independent), $k_B$ is the Boltzmann constant, *m* is a mass of $^{57}$Fe atoms.

The spectra measured on the Cr sample were analyzed in terms of sinusoidally modulated incommensurate spin-density waves (SDWs). The transferred hyperfine magnetic field seen by the $^{119}$Sn nuclei at crystal lattice position was assumed to be proportional to the SDW amplitude at this position. The first and the third harmonics as established in paper [2] were sufficient to describe the spectra. The spectra recorded on the Cr3%Mn sample were fitted using a hyperfine field distribution method as outlined in [11]. Both methods yielded distribution curves of the hyperfine field, p(B), and values of the center shifts, CS(T). Examples of the P(B) curves are displayed in Figs.2 and 3. Noteworthy is a difference in the shape of the p(B) curves. Dependences of CS on T are shown in Fig.4, and the solid lines represent the best-fit of the data to Eq.(1). Values of the Debye temperature, $T_D$, derived from these fits are as follows: $T_D$=477(56)K for Cr and $T_D$=282(16)K for Cr3%Mn. From the two figures it follows that the chromium lattice with the incommensurate SDWs is significantly harder than the one with a normal AF structure. The magnetism of the latter is much stronger than that of the former. As shown in Fig. 4, the average hyperfine field, $< B > = \int p(B) dB$, calculated for the Cr3%Mn sample is much larger than the one found for the Cr sample. Also the value of the Néel temperature for the Mn-doped Cr is about two-fold bigger. This means that it is not the strength of magnetism but rather its character (commensurate vs. incommensurate) that is responsible for the EPI, hence the difference in $T_D$.

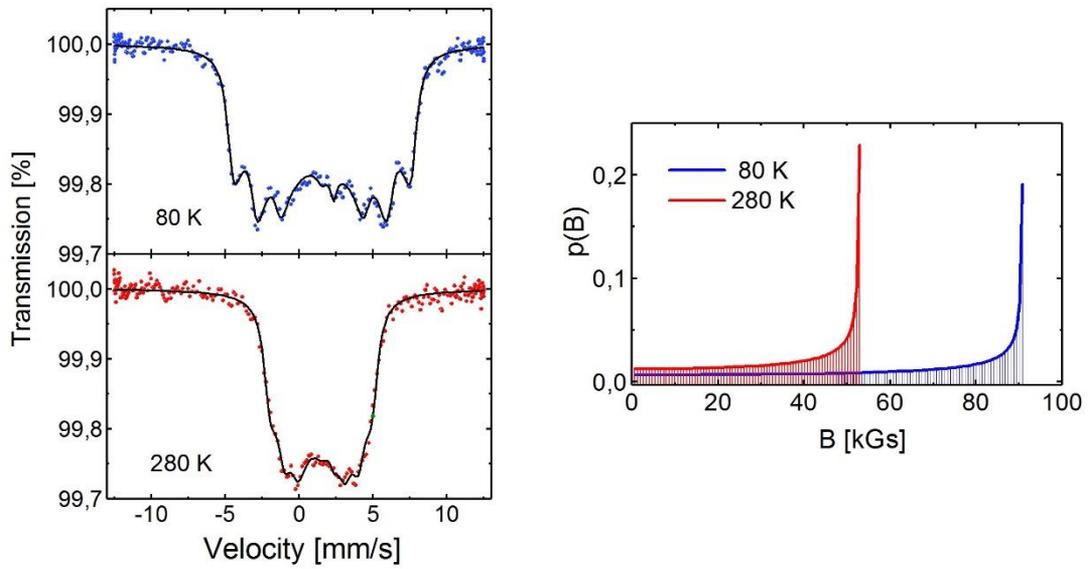

Fig. 2 Examples of the spectra and corresponding p(B)-curves for the Cr sample.

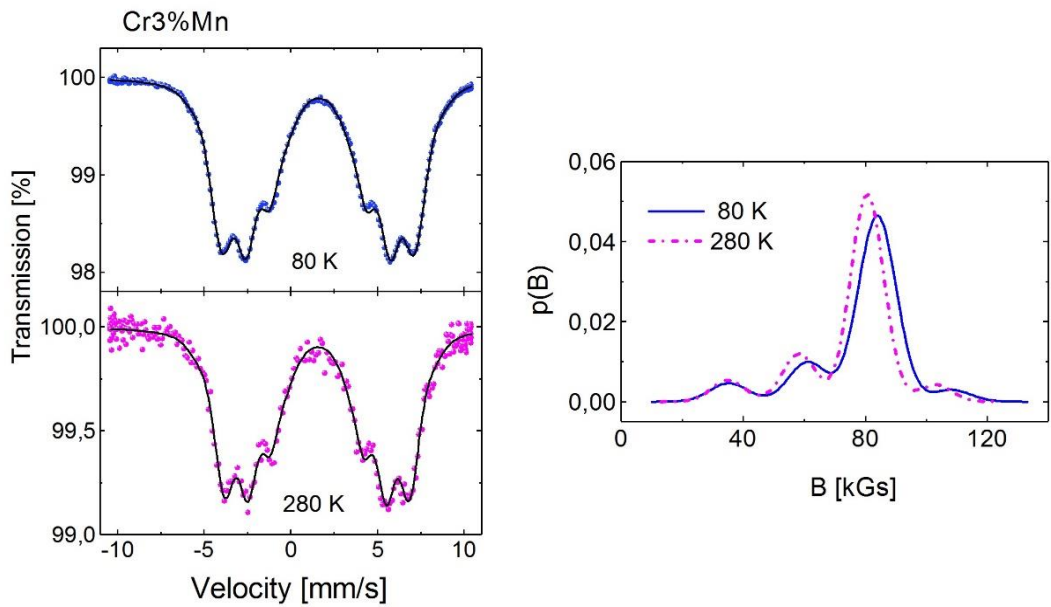

Fig. 3 Examples of the spectra and corresponding p(B)-curves for the Cr3%Mn sample.

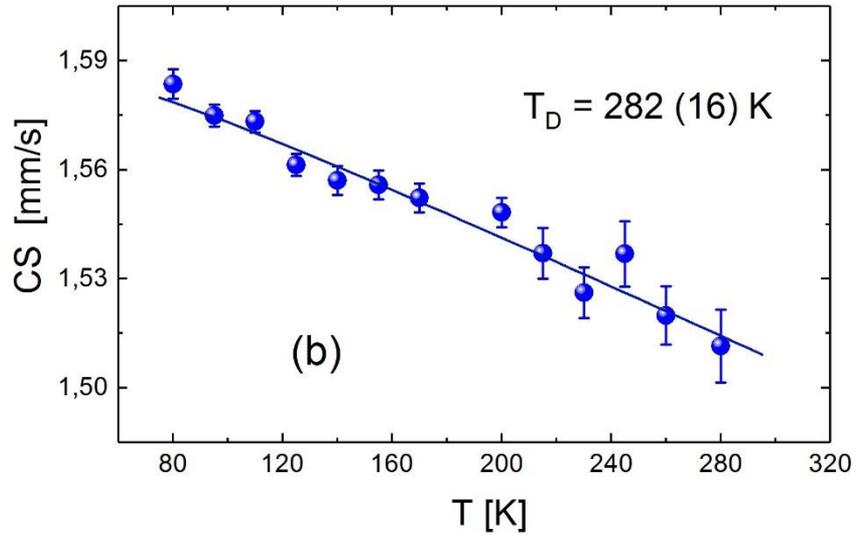

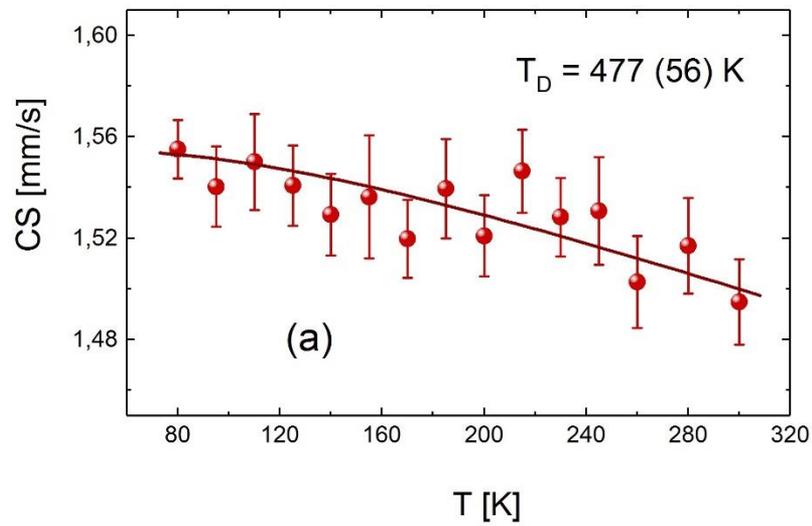

Fig. 4 Central shift, CS, vs. temperature, T, for (a) Cr and (b) Cr3%Mn. The solid lines stand for the best-fit of the data to Eq. (1). The values of the Debye temperature are displayed.

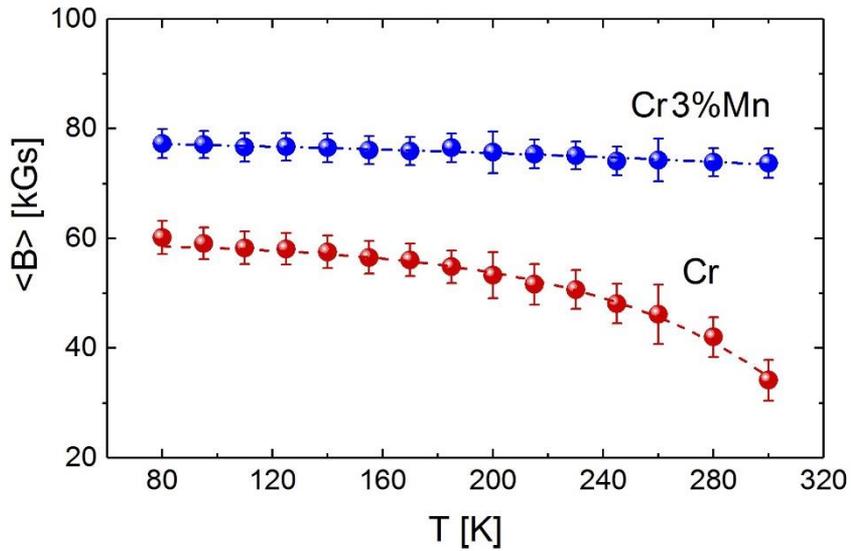

Fig. 5 Dependence of the average hyperfine field, <B>, on temperature, T, as determined for the studied samples. Dash lines are the guides to the eye.

In summary, the values of the Debye temperature, $T_D$, have been determined for single-crystals of a pure Cr, whose magnetic structure is constituted by incommensurate SDWs, and for Cr3%Mn having a normal (commensurate) AF structure. The value of $T_D$ determined for the pure Cr is by ~50% larger than the one found for the Cr3%Mn. As the magnetism of Cr is significantly weaker than the one of Cr3%Mn, the higher $T_D$-value found for the former can be regarded as evidence in favor of a stronger electron-phonon coupling in the incommensurate structure. In other words, the coupling is more affected by the type of magnetism rather than by its strength.

**Acknowledgements**

This work was financed by the Faculty of Physics and Applied Computer Science AGH UST and ACMIN AGH UST statutory tasks within subsidy of Ministry of Science and Higher Education, Warszawa. The single-crystals were donated by Eric Fawcett.